\documentstyle[preprint,aps,pra,epsfig]{revtex}
\tighten
\begin{document}
\draft
\widetext
\preprint{version 2.0 }

\title{Hyperfine splitting of $2^3S_1$ state in He$^3$}

\author{Krzysztof Pachucki
\thanks{E-mail address: krp@fuw.edu.pl}}
\address{
Institute of Theoretical Physics, Warsaw University,
Ho\.{z}a 69, 00-681 Warsaw, Poland}
\maketitle

\begin{abstract}
Relativistic corrections to the hyperfine splitting are calculated for 
the triplet $2^3S_1$ state of the helium isotope He$^3$.
High precision variational wave functions are employed, where
the electron-electron correlations are well accounted for.
Due to the unknown nuclear structure a comparison with
the experimental result is performed via He$^{3+}$ hyperfine splitting.
Surprisingly, the second order contribution due to Fermi interaction, in spite
of the additional small factor $m/M$, is very significant.

\end{abstract}
\narrowtext

The calculation of higher order relativistic and QED effects in 
few electron systems is a long standing problem. Various measurements
of transition frequencies have reached the precision of few ppb,
while no theoretical predictions are yet so accurate \cite{drake}.
The usual approaches, which incorporate most of relativistic effects from 
the beginning, MCDF for example, are not capable to include  
electron-electron correlations in a complete way,
thus their predictions are not sufficiently accurate.  
The alternative approach starts from the Schr\"odinger
equation and incorporates relativistic effects perturbatively in the
effective hamiltonian. The principal advantage is the simplicity 
and high accuracy of nonrelativistic wave functions, 
which allows for precise calculations of higher order relativistic effects. 
The cost one pays in the perturbative approach
is the complexity and singularity of the effective hamiltonian.
Historically, the helium fine structure splitting 
was the first application of the higher order  effective (Breit-like) 
hamiltonian. Douglass and Kroll in \cite{doug} with the help of Bethe-Salpeter 
formalism derived effective operators at order $m\,\alpha^6$, which
together with the second order correction from  leading Breit terms
gave a {\em complete} $m\,\alpha^6$ contribution to helium fine structure
of $2P_J$ levels. It was the first disprove of the common believe, 
that Breit Hamiltonian could not be used at higher orders of perturbation calculus.
20 years later Khriplovich and coworkers \cite{khr} essentially 
rederived their result, without being aware of this former work,
using the time ordered perturbation technique. With these effective operators
they calculated fine structure splitting in positronium. The extension
of the perturbative approach to S-states is a highly nontrivial task, see 
for example \cite{krp2}. It is due to high singularity of effective operators, 
which matrix elements between S-states diverge, so one may question
about the correctness of the perturbative approach. 
In spite of these divergences, 
it is possible to implement the perturbative approach to S-states as well.
The idea lies in the regularization of Coulomb interaction, in such a way
that all matrix elements become finite. All contribution coming from 
high-momenta are incorporated as local interaction, proportional to Dirac delta
or it's derivatives. The coefficients are found by matching
for example, the scattering amplitude and they depend on 
the regularization parameter. However, this dependence cancels out
when all terms contributing to energy at specified order are summed up. 
As a test, one may rederive \cite{krp2} Dirac energy levels at order $m\,\alpha^6$.
This approach was successfully applied first in the calculation
of positronium hfs \cite{krp2}. This result was confirmed independently 
by 3 groups \cite{conf},
two of them were applying directly Bethe-Salpeter equation, and this really 
serves as an independent check of this method. Little later, one performed \cite{krp1}
calculations of higher order relativistic corrections to helium $2^3S_1$ level.
The fact, that it is a triplet state, was a great simplification, since the 
wave function vanishes at ${\bf r}_1={\bf r}_2$. Very recently,
Yelkhovsky \cite{yelkh} has derived complete set of operators for the ground
$1^1S_0$ state. The numerical calculation of their matrix elements is still under way. 
In this work we study He$^3$ hyperfine splitting,
which was measured very precisely by Rosner and Pipkin in \cite{hfs}.
He$^3$ hfs is a nice example of difficulties 
in the bound state QED and perturbative approaches. The Fermi interaction, 
as given by Dirac $\delta^3(r)$ function is already quite singular. 
Incorporation of further relativistic effects leads to even more singular 
operators, which have to be properly handled. 
In the work  \cite{krp1} we have shown the practical way 
to handle singular effective operators for helium atom.
Coulomb interaction is smoothed at small distances 
with the use of some parameter $\lambda$, to avoid small $r$ nonintegrable 
singularities. The key point of this approach lies in the fact, 
that dependence on $\lambda$ cancel out between all matrix elements, 
and  correct energy levels are restored in the limit $\lambda\rightarrow\infty$.
High electron momenta does not contribute here, because it is again triplet S-state,
so one does not need to construct counter terms for matching scattering amplitude.
Surprizingly, we have found a very large second order correction due to
Fermi interaction. This correction requires a separate treatment
and will be dealt with at the end of this work.

Hyperfine splitting in He$^3$ of $2^3S_1$ state is due to the interaction
of electron and helion (nucleus) magnetic moments. In the $2^3S_1$ state
both electron spins are parallel and sum to $S=1$ in contrast to the ground state
were $S=0$. Magnetic moment of helion comes mainly from the neutron particle.
It means that helion g-factor is negative and hyperfine sublevels are inverted
with respect to hydrogen, the upper one has $S+I=1/2$ and the lower one $S+I=3/2$.
Therefore, by a hyperfine splitting one means here $E_{\rm hfs}= E(1/2)-E(3/2)$.  
According to the perturbative  approach the general expression for the 
hyperfine splitting up to the order $m\,\alpha^6$ is:
\begin{equation}
E_{\rm hfs} = \langle H^{(4)}_{\rm hfs}\rangle
             +\langle H^{(5)}_{\rm hfs}\rangle+
              \langle H^{(6)}_{\rm hfs}\rangle+
2\,\langle H^{(4)}\,\frac{1}{(E-H)'}\,H^{(4)}_{\rm hfs}\rangle+E_{\rm rec}\,. \label{1}
\end{equation}
$H^{(4)}_{\rm hfs}$ is:
\begin{eqnarray}
H^{(4)}_{\rm hfs} &=& H^{A}_{\rm hfs}+H^{B}_{\rm hfs}+H^{D}_{\rm hfs}\,,\\
H^{A}_{\rm hfs} &=&  \frac{8\,Z\,\alpha}{3\,m\,M}\,\biggl[
\frac{\bbox{\sigma}_n\cdot\bbox{\sigma}_1}{4}\,\pi\,\delta^3(r_1)+
\{1\rightarrow 2\}\biggr](1+k)\,\,(1+a)\,,\label{3} \\
H^{B}_{\rm hfs} &=& \frac{Z\,\alpha}{2\,m\,M}\biggl[
\frac{\bbox{r}_1\times\bbox{p}_1}{r_1^3}+
\frac{\bbox{r}_2\times\bbox{p}_2}{r_2^3}\biggr]\cdot\bbox{\sigma}_n\,(1+k)\,,\\
H^{D}_{\rm hfs} &=& -\frac{Z\,\alpha}{4\,m\,M}\,\biggl[
\frac{\sigma_n^i\,\sigma_1^j}{r_1^3}\,
\biggl(\delta^{ij}-3\,\frac{r_1^i\,r_1^j}{r_1^2}\biggr)+\{1\rightarrow 2\}\biggr]\,(1+k)\,,
\end{eqnarray}
where $a,k$ are anomalous magnetic moments of the electron and the helion respectively.
However, it is not commonly accepted such a notion for a nucleus.
The relation of $k$ with the magnetic moment of the nucleus with charge $Z\,e$ is 
$\bbox{\mu}= 2\,(1+k)\,Z\,e/(2\,M)\,\bbox{I}$.
Masses $m$ and $M$ are of the electron and helion respectively. 
The expectation values
of $H^{B}_{\rm hfs}$ and $H^{D}_{\rm hfs}$ vanish in $2^3S_1$ state, but
they contribute in the second order, last term in Eq. (\ref{1}).
$H^{(4)}$ is a Breit hamiltonian in the nonrecoil limit:
\begin{eqnarray}
H^{(4)} &=& H^{A}+H^{B}+H^{D}\,,\label{6}\\
H^{A} &=& -\frac{1}{8\,m^3}\,(p_1^4+p_2^4)+
\frac{Z\,\alpha\,\pi}{2\,m^2}\,\bigl[\delta^3(r_1)+\delta^3(r_2)\bigr]-
\frac{\alpha}{2\,m^2}\,p_1^i\biggl(
\frac{\delta^{ij}}{r}+\frac{r^i\,r^j}{r^3}\biggr)\,p_2^j\,, \\
H^{B} &=& \biggl[\frac{Z\,\alpha}{4\,m^2}\biggl(
\frac{\bbox{r}_1\times\bbox{p}_1}{r_1^3}+
\frac{\bbox{r}_2\times\bbox{p}_2}{r_2^3}\biggr)-
\frac{3\,\alpha}{4\,m^2}\,\frac{\bbox{r}}{r^3}\times(\bbox{p}_1-\bbox{p}_2)\biggr]\,
\frac{\bbox{\sigma}_1+\bbox{\sigma}_2}{2}\,,\\
H^{D} &=& \frac{\alpha}{4\,m^2}\,
\frac{\sigma_1^i\,\sigma_2^j}{r^3}\biggl(
\delta^{ij}-\frac{3\,r^i\,r^j}{r^2}\biggr)\,,
\end{eqnarray}
where $\bbox{r} = \bbox{r}_1-\bbox{r}_2$ and $r=|\bbox{r}|$.
Any possible recoil corrections at order $m\,\alpha^6$ are included in $E_{\rm rec}$. 
In spite of the small additional factor $m/M$ their contribution 
is significant. This effect \cite{stern} could be associated to the mixing 
of $2^3S_1$ and $2^1S_1$ states caused by the Fermi interaction.
$\langle H^{(5)}_{\rm hfs}\rangle$ is delta-like term with the coefficient
given by the two-photon forward scattering amplitude. It is the same 
like in hydrogen and strongly depends on the nuclear structure. It also 
automatically includes nuclear recoil effects 
and inelastic contribution (nuclear polarizability).
$\langle H^{(5)}_{\rm hfs}\rangle$ has been considered in detail for 
the case of hydrogen and muonic hydrogen.
However, there are no sufficient experimental data available for helion,
He$^3$ nucleus, therefore we were unable to estimate these contributions.
Moreover, it would be incorrect to apply this correction for point-like nucleus,
because high energy photon momenta are involved,
where nucleus, could not be approximated as a point like particle.
Therefore we will leave this contribution unevaluated, and at the end
for the comparison with an experiment,
will subtract the appropriately scaled hydrogenic value for hyperfine splitting   
The last term  $H^{(6)}_{\rm hfs}$ includes spin dependent operators,
which contribute at order $m\,\alpha^6$. It is only this term, 
which derivation was not performed so far in the literature 
and is presented here. The detailed description on derivation of
effective hamiltonian, reader may find in former works,
for example in \cite{krp2}. There are four time ordered diagrams, which contributes
to $H^{(6)}_{\rm hfs}$ and are presented in Fig~ 1. The first two
are the same as in hydrogen, other two are essentially three body terms. 
One derives the following expressions, which corresponds to these diagrams.
\begin{eqnarray}
H^{(6)}_{\rm hfs} &=& V_1+V_2+V_3+V_4\,,\\
V_1 &=&-(1+k)\,\,\frac{\bbox{\sigma}_n\cdot\bbox{\sigma}_1}{48\,M\,m^3}\,
\biggl\{
2\,p_1^2\,4\,\pi\,Z\,\alpha\,\delta^3(r_1)+
2\cdot4\,\pi\,Z\,\alpha\,\delta^3(r_1)\,p_1^2+
\biggl[p_1^2,\biggr[p_1^2,\frac{Z\,\alpha}{r_1}\biggr]\biggr]\biggr\}
\nonumber \\ &&+\{1\rightarrow2\}\,,\\
V_2 &=& (1+k)\,\,\frac{(Z\,\alpha)^2}{r_1^4}\,
\frac{\bbox{\sigma}_n\cdot\bbox{\sigma}_1}{6\,M\,m^2}+\{1\rightarrow2\}\,,\\
V_3 &=& -(1+k)\,\,\frac{\bbox{\sigma}_n\cdot\bbox{\sigma}_1}{6\,M\,m^2}\,
\frac{Z\,\alpha\,\bbox{r}_1}{r_1^3}\cdot\frac{\alpha\,\bbox{r}}{r^3}+
\{1\leftrightarrow 2\}\,,\\
V_4 &=& -(1+k)\,\,\frac{\bbox{\sigma}_n\cdot\bbox{\sigma}_2}{6\,M\,m^2}\,
\frac{Z\,\alpha\,\bbox{r}_1}{r_1^3}\cdot\frac{\alpha\,\bbox{r}}{r^3}+
\{1\leftrightarrow 2\}\,. 
\end{eqnarray} 
It is worth noting that $V_1$ ans $V_2$ are the same as in hydrogen.
Matrix element of $H^{(6)}_{\rm hfs}$ and last term in Eq. (\ref{1}) are 
separately divergent at small $r_1$ and $r_2$. However these divergences cancel out 
in the sum. It is because the high electron momentum contribution
is absent in the nonrecoil limit. For hydrogen this sum is equal to
the  expectation value of $\gamma\,A$ on Dirac wave function. We introduce now 
the following regulator $\lambda$ to the electron-nucleus Coulomb interaction
\begin{equation}
\frac{Z\,\alpha}{r_i}\rightarrow \frac{Z\,\alpha}{r_i}\,
\bigl(1-e^{-\lambda\,m\,Z\,\alpha\,r_i}\bigr)\,,
\end{equation}
in all hamiltonians in Eq.(\ref{1}), as well as in the nonrelativistic one.
This leads to the following further replacements in $H^{(6)}_{\rm hfs}$
\begin{eqnarray}
4\,\pi\,Z\,\alpha\,\delta^3(r_i) &\equiv& -\nabla^2\,\frac{Z\,\alpha}{r_i}
\rightarrow -\nabla^2\,\frac{Z\,\alpha}{r_i}
\bigl(1-e^{-\lambda\,m\,Z\,\alpha\,r_i}\bigr)\,, \\ 
\frac{(Z\,\alpha)^2}{r_i^4} &\equiv& 
\biggl(\nabla\frac{Z\,\alpha}{r_i}\biggr)^2 \rightarrow
\biggl(\nabla\frac{Z\,\alpha}{r_i}
\bigl(1-e^{-\lambda\,m\,Z\,\alpha\,r_i}\bigr)\biggr)^2\,.
\end{eqnarray}
Once the interaction is regularized, one can calculate all matrix elements
and take the limit $\lambda\rightarrow\infty$. As a first step,
using formulas from \cite{krp2}, we rederived the known relativistic 
correction to hfs in hydrogen
\begin{equation}
\delta E_{\rm hfs} = (1+k)\,\frac{\mu^3}{m\,M}\,
\frac{(Z\,\alpha)^6}{n^3}\,
\frac{\bbox{\sigma}_n{\bbox{\sigma}_e}}{4}\,\biggl(
\frac{44}{9}+\frac{4}{n}-\frac{44}{9\,n^2}\biggr)\,, \label{18}
\end{equation}
where $n$ is a principal quantum number. It agrees with that, 
obtained directly from the Dirac equation. Since for helium, all matrix elements 
could be calculated only numerically, we will transform effective operators
to the regular form, where $\lambda$ could be taken to infinity
before the numerical calculations. The initial expression
for a complete set of relativistic corrections in atomic units is
(with implicit $\lambda$ regularization and excluding recoil):
\begin{eqnarray}
\delta E_{\rm hfs} &=& |1+k|\,\,\frac{\mu^3}{m\,M}\,\alpha^6\,{\cal E}\,,\\
{\cal E} &=& {\cal E}_A+{\cal E}_B+{\cal E}_D+{\cal E}_N\,,\\
{\cal E}_A &=& 2\biggl\langle\biggl\{
-\frac{1}{8}\,(p_1^4+p_2^4)+
\frac{Z\,\pi}{2}\,\bigl[\delta^3(r_1)+\delta^3(r_2)\bigr]-
\frac{1}{2}\,p_1^i\,\biggl(\frac{\delta^{ij}}{r}+\frac{r^i\,r^j}{r^3}\biggr)
\biggr\}\nonumber \\ &&\frac{1}{(E-H)'}\,
2\,Z\,\pi\,\bigl[\delta^3(r_1)+\delta^3(r_2)\bigr]\biggr\rangle\,,\\ 
\nonumber \\
{\cal E}_B &=& 2\biggl\langle\biggl\{\frac{Z}{4}\biggl[
\frac{\bbox{r}_1\times\bbox{p}_1}{r_1^3}+
\frac{\bbox{r}_2\times\bbox{p}_2}{r_2^3}\biggr]
-\frac{3}{4}\,\frac{\bbox{r}\times(\bbox{p}_1-\bbox{p}_2)}{r^3}\biggr\}
\frac{1}{(E-H)'}\,\nonumber \\ &&
\frac{Z}{2}\,\biggl[
\frac{\bbox{r}_1\times\bbox{p}_1}{r_1^3}+
\frac{\bbox{r}_2\times\bbox{p}_2}{r_2^3}\biggr]\biggr\rangle\,,
\\
{\cal E}_D &=& 2\biggl\langle\frac{1}{4}\,\biggl(
\frac{\delta^{ij}}{r^3}-3\,\frac{r^i\,r^j}{r^5}\biggr)\,\frac{1}{(E-H)'}\,
\biggl(-\frac{Z}{4}\biggr)\biggl[
\biggl(\frac{\delta^{ij}}{r_1^3}-3\,\frac{r_1^i\,r_1^j}{r_1^5}\biggr)+
\biggl(\frac{\delta^{ij}}{r_2^3}-3\,\frac{r_2^i\,r_2^j}{r_2^5}\biggr)
\biggr]\biggr\rangle\,,\\
\nonumber \\
{\cal E}_N &=& \biggl\langle
-\frac{1}{4}\,\biggl(p_1^2\,\delta^3(r_1)+p_2^2\,\delta^3(r_2)\biggr)
-\frac{1}{16}\,\biggl(\biggl[p_1^2,\biggl[p_1^2,\frac{Z}{r_1}\biggr]\biggr]
 + \biggl[p_2^2,\biggl[p_2^2,\frac{Z}{r_2}\biggr]\biggr]\biggr)
\nonumber \\ && 
+\frac{1}{2}\,\biggl(\frac{Z^2}{r_1^4}+\frac{Z^2}{r_2^4}\biggr)
-\frac{\bbox{r}}{r^3}\,\biggl(\frac{Z\,\bbox{r}_1}{r_1^3}-\frac{Z\,\bbox{r}_2}{r_2^3}\biggr)
\biggr\rangle\,,
\end{eqnarray}
where ${\cal E}_N=\langle H^{(6)}_{\rm hfs}\rangle$, and we used the following 
formulas for hfs of $^3S_1$ states:
\begin{eqnarray} 
\langle\bbox{\sigma}_n\cdot\bbox{\sigma}_1\rangle = 
\langle\bbox{\sigma}_n\cdot(\bbox{\sigma}_1+\bbox{\sigma}_2)/2\rangle &=&-3\,, \\ 
\langle\sigma_1^i\,\sigma_2^j\,Q_1^{ij}\;
\sigma_n^a\,(\sigma_1+\sigma_2)^b\,Q_2^{ab}\rangle &=&
-2\,Q_1^{ij}\,Q_2^{ij}\,,
\end{eqnarray}
for symmetric and traceless $Q^{ij}$.  There is also a one loop radiative correction
\begin{equation}
{\cal E}_R = 2\,Z^2\,\biggl(\ln 2 - \frac{5}{2}\biggr)\,
\langle\pi\,\delta^3(r_1)+\pi\,\delta^3(r_2)\rangle\,,
\end{equation} 
which is similar to that in hydrogen. It will not contribute to the
special difference between the helium and hydrogen-like helium hfs,
therefore we will not consider it any further.
The initial expression is rewritten to the regular form,
where $\lambda$ regularization is not necessary. 
The operators in second order terms ${\cal E}_A$ 
are transformed with the use of
\begin{eqnarray}
H'^A &\equiv& H^A-
\frac{1}{4}\,\biggl(\frac{Z}{r_1}+\frac{Z}{r_2}\biggr)\,(E-H)-
\frac{1}{4}\,(E-H)\,\biggl(\frac{Z}{r_1}+\frac{Z}{r_2}\biggr)\,, \\
4\,\pi\,Z\,[\delta^3(r_1)+\delta^3(r_2)]' &\equiv&
4\,\pi\,Z\,[\delta^3(r_1)+\delta^3(r_2)]
\nonumber \\ &&+
2\,\biggl(\frac{Z}{r_1}+\frac{Z}{r_2}\biggr)\,(E-H)+
2\,(E-H)\,\biggl(\frac{Z}{r_1}+\frac{Z}{r_2}\biggr)\,. \label{29} 
\end{eqnarray}
This transformation leads to new form for ${\cal E}'_A$ and ${\cal E}'_N$,
such that
\begin{eqnarray}
{\cal E}_A+{\cal E}_N &=& {\cal E}'_A+{\cal E}'_N \,,\\
{\cal E}'_A &=& 2\,\biggl\langle H'^A\frac{1}{(E-H)'}\,
2\,Z\,\pi\,\bigl[\delta^3(r_1)+\delta^3(r_2)\bigr]'\biggr\rangle\,,\\ 
\nonumber \\
{\cal E}'_N &=& \biggl\langle \biggl(E-\frac{1}{r}\biggr)^2\,
\biggl(\frac{Z}{r_1}+\frac{Z}{r_2}\biggr)+
\biggl(E-\frac{1}{r}\biggr)\,
\biggl(\frac{Z^2}{r_1^2}+\frac{Z^2}{r_2^2}+
4\,\frac{Z}{r_1}\,\frac{Z}{r_2}\biggr)+
2\,\frac{Z}{r_1}\,\frac{Z}{r_2}\,\biggl(\frac{Z}{r_1}+\frac{Z}{r_2}\biggr)
\nonumber \\ &&
-\biggl(E-\frac{1}{r}+\frac{Z}{r_2}-
\frac{p_2^2}{2}\biggr)\,4\,\pi\,Z\,\delta^3(r_1)-
\frac{5\,Z}{4}\,\frac{r^i}{r^3}\,\biggl(
\frac{r^i_1}{r_1^3}-\frac{r^i_2}{r_2^3}\biggr)
\nonumber \\ &&
+p^i_1\,\frac{Z^2}{r_1^2}\,p^i_1
-p_2^2\,\frac{Z}{r_1}\,p_1^2
+2\,p_2^i\,\frac{Z}{r_1}\,\biggl(
\frac{\delta^{ij}}{r}+\frac{r^i\,r^j}{r^3}\biggr)\,p_1^j\biggr\rangle
\nonumber \\ &&
-\frac{1}{4}\,\biggl\langle\frac{Z}{r_1}+\frac{Z}{r_2}\biggr\rangle\,
\langle 4\,\pi\,Z\,(\delta^3(r_1)+\delta^3(r_2))\rangle
+2\,\biggl\langle\frac{Z}{r_1}+\frac{Z}{r_2}\biggr\rangle\,
\langle H^A\rangle\,.
\end{eqnarray}
In the numerical calculations of these matrix elements 
we follow the approach developed by Korobov \cite{kor}.
The $S$ wave function is expanded in the sum of pure exponentials
\begin{equation}
\phi = \sum_{i=1}^N v_i\,\bigl(
e^{-\alpha_i\,r_1-\beta_i\,r_2-\gamma_i\,r}-(r_1\leftrightarrow r_2)\bigr)\,,
\end{equation}
with randomly chosen $\alpha_i,\beta_i,\gamma_i$ in some specified limits.
This basis set has been proven to give excellent results
for the nonrelativistic energy and the wave function. Moreover, its simplicity
allows for the calculations of relativistic corrections.
With basis set $N=1200$ we obtained the nonrelativistic energy 
(without the mass polarization term $\bbox{p}_1\bbox{p}_2\,\mu/M$)
\begin{equation}
E=-2.1752293782367913057(1)\,,
\end{equation}
slightly below the previous result in \cite{Yan}.
Expectation values of Dirac delta function without 
and with the mass polarization term are correspondingly:
\begin{eqnarray}
\langle\,4\,\pi\,(\delta^3(r_1)+\delta^3(r_2))\rangle &=&33.184142630(1)\,, \\
\langle\,4\,\pi\,(\delta^3(r_1)+\delta^3(r_2))\rangle_{\rm MP} &=&33.184152589(1)\,.
\end{eqnarray}
The last one gives the leading hfs in helium, which is
\begin{equation}
E_{\rm hfs} = 2\,Z\,\alpha^4\,\frac{\mu^3}{m\,M}\,|1+k|\,(1+a)\,
\langle\pi\,(\delta^3(r_1)+\delta^3(r_2))\rangle_{\rm MP} 
\approx 6\,740\,451\,{\rm kHz}\,,
\label{37}
\end{equation}
where we use values of physical constants from Ref. \cite{mohr}
with one exception \cite{exception}.
Numerical results for ${\cal E}_X$ with $X=A,B,D,N$ are presented 
in Table I. The inversion of $H-E$ in ${\cal E}_A$ is performed in 
the similar basis set as
for $2^3S_1$ wave function, however the nonlinear parameters had have
to be properly chosen, to obtain a sufficiently accurate result.
Namely, if $0<X,Y,Z<1$ are independent pseudo random numbers with homogeneous
distribution, then:
\begin{eqnarray}
\alpha &=& A_2\,X^{-n}+A_1\,,\\
\beta  &=& (B_2-B_1)\,Y+B_1\,,\\
\gamma &=& (C_2-C_1)\,Z+C_1\,.
\end{eqnarray}
Parameters $A,B,C$ and $n$ are found, by minimization 
of the second order term with regularized Dirac delta on both sides. 
The inversion of $H-E$ in ${\cal E}_B$ is performed in 
the basis set of the form
\begin{equation}
\bbox{\phi} = \bbox{r}_1\times\bbox{r}_2\,\sum_{i=1}^N v_i\,\bigl(
e^{-\alpha_i\,r_1-\beta_i\,r_2-\gamma_i\,r}+(r_1\leftrightarrow r_2)\bigr)\,.
\end{equation}
Unfortunately, we have not been able to get a reliable number
for ${\cal E}_D$ with this numerical approach. The reason is 
that operators in ${\cal E}_D$ are so singular, that this basis set gives 
a very pure convergence. The result presented in Table I, is obtained
analytically within $1/Z$ approximation, namely we neglected
completely electron-electron interaction and corrected this value
by factor $\langle\pi(\delta^3(r_1)+\delta^3(r_2))\rangle/9$.
The estimated uncertainty is of the order of 10\%. 
The total contribution of $m\,\alpha^6$ term (without recoil $E_{\rm rec})$ to hfs is
\begin{equation}
E^{(6)} = |1+k|\,\,\frac{\mu^3}{m\,M}\,\alpha^6\,200.9844(5) = 2171.440(5)\,{\rm kHz}\,, 
\end{equation}
what could compared to the leading Fermi contact interaction in Eq. (\ref{37}),
$E^{(6)}/E_{\rm hfs} \approx 0.000322$.
Between all the contributions to He$^3(2^3S_1)$ hyperfine splitting in Eq.(\ref{1}),
$H^{(5)}$, essentially the nuclear structure contribution, requires
input from the nuclear physics to be reliable evaluated. This is the reason
we do not present final theoretical predictions for hfs, to  compare with 
the precise measurement in \cite{hfs} 
\begin{equation}
E_{\rm hfs}({\rm He}) = 6\,739701.177(16)\,{\rm kHz}\,.
\end{equation}
Instead, we can compare our result indirectly by subtracting the ground state 
hfs of helium ion as measured in \cite{hfs+}
\begin{equation}
E_{\rm hfs}({\rm He}^+) = 8\,665\,649.867(10)\,{\rm kHz}\,,
\end{equation}
by composing the following difference
\begin{equation}
\Delta E_{exp} = E_{\rm hfs}({\rm He})-\frac{3}{4}\,
\frac{\langle \pi\,(\delta^3(r_1)+\delta^3(r_2))\rangle_{\rm MP}}{8}\,
E_{\rm hfs}({\rm He}^+) = -38.998(19)\,{\rm kHz}\,. \label{45}
\end{equation}
In this way, nuclear structure contribution, of order $m\,\alpha^5$
cancels out, as well as the leading Fermi contact interaction.
What remain are electron--electron correlation effects.
Theoretical predictions for this difference are (see Table I)
\begin{equation}
\Delta E_{rel} = |1+k|\,\,\frac{\mu^3}{m\,M}\,\alpha^6\,1.8795(5)
 =20.306(5)\,{\rm kHz}\,. \label{46}
\end{equation}
We do not associate here the uncertainty due to higher order terms.
A disagreement of the theoretical result in Eq. (\ref{46}) with the experimental 
one in Eq. (\ref{45}) indicates that the presented calculation is incomplete.
However the old calculations of relativistic corrections in \cite{dhf} using 
an approximate DHF wave function, led to the result which is in an 
agreement with experiment $ \Delta E_{\rm old} = -32(22)\,{\rm kHz}$.
This agreement is only  illusive, since there is a correction
discovered by Sternheim \cite{stern}, which is denoted here by $E_{\rm rec}$. 
It is the second order contribution due to Fermi interaction $H^A_{\rm hfs}$ 
in Eq. (\ref{3}) 
\begin{equation}
E_{\rm rec} = \langle\,H^A_{\rm hfs}\,\frac{1}{(E-H)'}\,H^A_{\rm hfs}\,\rangle\,.
\label{47}
\end{equation}
It could be understood as a recoil correction since it 
includes additional small factor $m/M$. Sternheim noticed that $H^A_{\rm hfs}$
mixes $2^3S_1$ and $2^1S_1$ states, what together with small energy difference
between these states leads to strong enhancement of this recoil correction.
Sternheim result which includes only $2^1S_1$ intermediate state is $\delta E=-66.7(3)$ kHz,
what nicely would explain the discrepancy of 59.3 kHz. However, the inclusion
of higher excited states, will lead to the infinite result. Moreover, 
this correction is partially included in $\langle H^{(5)}_{\rm hfs}\rangle$,
and only after proper subtraction it becomes finite. Hopefully, all divergence
or in other words dependence on regularization parameter $\lambda$ cancel out
in this particular difference between helium atom and helium ion
hyperfine splitting. Therefore instead of Eq. (\ref{47}) we calculate the difference
\begin{eqnarray}
\Delta E_{\rm rec} &=& E_{\rm rec}({\rm He})-\frac{3}{4}\,
\frac{\langle\pi\,(\delta^3(r_1)+\delta^3(r_2))\rangle_{\rm He}}{8}\,E_{\rm rec}({\rm He}^+)
= \frac{3}{2}\,\biggl(\frac{Z\,\alpha\,(1+k)}{3\,m\,M}\biggr)^2 \nonumber \\ &&
\times\biggl\{ \frac{1}{2}\,
\biggl\langle 4\,\pi\,[\delta^3(r_1)+\delta^3(r_2)]\,\frac{1}{(E-H)'}\,
 4\,\pi\,[\delta^3(r_1)+\delta^3(r_2)]\biggr\rangle \nonumber \\ &&
+\frac{1}{2}\,
\biggl\langle 4\,\pi\,[\delta^3(r_1)-\delta^3(r_2)]\,\frac{1}{E-H}\,
 4\,\pi\,[\delta^3(r_1)-\delta^3(r_2)]\biggr\rangle\nonumber \\ &&
-\frac{\langle\pi\,(\delta^3(r_1)+\delta^3(r_2))\rangle_{\rm He}}{8}\,
\biggl\langle 4\,\pi\,\delta^3(r)\,\frac{1}{(E-H)'}\,
 4\,\pi\,\delta^3(r)\biggr\rangle_{{\rm He}^+}\biggr\}
\end{eqnarray}
Further evaluations proceed in the similar way as for ${\cal E}_A$.
With the help of Eq.(\ref{29}) one subtracts algebraically all divergences
and the remaining finite terms are calculated numerically.
The result is
\begin{equation}
\Delta E_{\rm rec} = -\frac{3}{2}\,\biggl(\frac{Z\,\alpha\,(1+k)}{3\,m\,M}\biggr)^2\,
                    m^5\,\alpha^4\,14491.0(1) 
                   = - |1+k|\,\,\frac{m^2}{M}\,\alpha^6\, 5.5971 
                   = -60.471\,{\rm kHz} \label{49}
\end{equation}
The complete theoretical predictions are the sum of Eqs. (\ref{46}) and (\ref{49}) 
\begin{equation}
\Delta E_{\rm th} = - |1+k|\,\,\frac{\mu^3}{m\,M}\,\alpha^6\,3.71756 = -40.165\,{\rm kHz}
\end{equation}
It nicely agrees with the experimental value --38.998 kHz from Eq. (\ref{45}). 
The small difference of 1.167 kHz
could be associated to higher order QED corrections. 
In summary we have calculated relativistic and a dominating recoil correction to 
helium hyperfine splitting.
Due to unknown nuclear structure we considered such a difference of hfs of
helium atom and He$^+$ ion, that cancels out this nuclear structure contribution.
The recoil correction, namely the second order term in the Fermi interaction
give a significant contribution to hfs, due to mixing of singlet and triplet S-states. 
Was this correction included, a good agreement with experiment is achieved.
 
\section*{ACKNOWLEDGMENTS}
I gratefully acknowledge helpful information about experimental results from Peter Mohr.
This work was supported by Polish Committee for Scientific Research 
under Contract No. 2P03B 057 18.

\begin{figure}
\centerline{\epsfig{figure=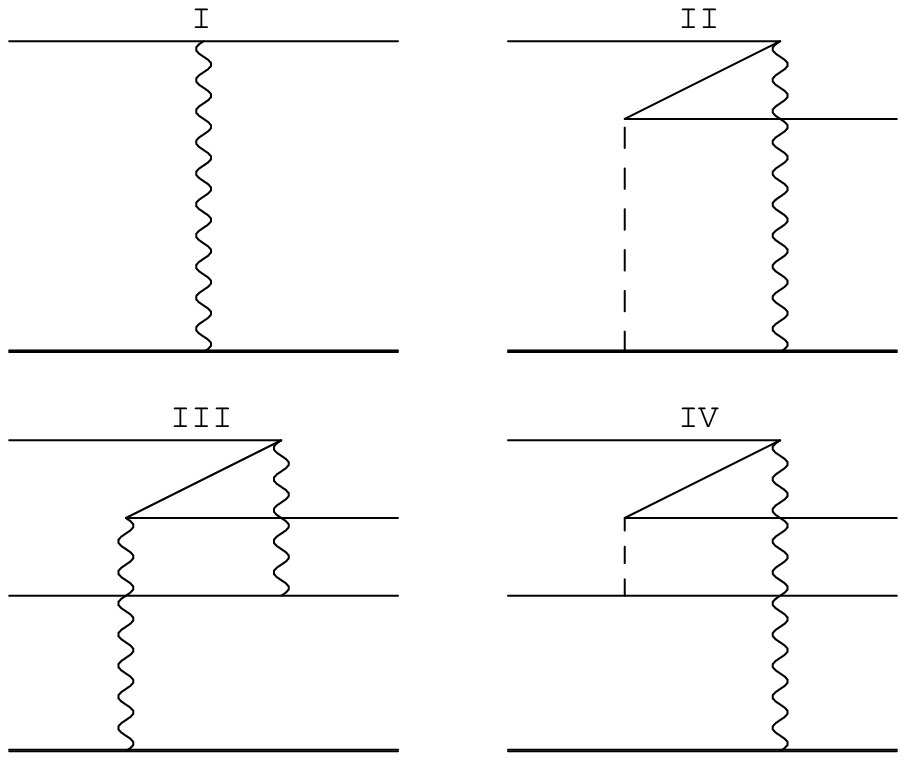,width=4in}}
 \caption{Time ordered diagrams contributing to helium 
hyperfine structure at order $m\,\alpha^6$.
Dashed line is a Coulomb photon, the wavy line is the transverse photon,
the thicker vertical line denotes nucleus, two other electrons. }
\label{fig1}
\end{figure}

\begin{table}
\begin{center}
\begin{tabular}{|l|r|} 
contribution			&$|1+k|\,\,m^2/M\,\alpha^6$\\ \hline
${\cal E}_A$			&202.6761\hspace{4ex} \\ 
${\cal E}_B$			&  0.0059\hspace{4ex} \\
${\cal E}_D$			&  0.0054(5)\hspace{1ex} \\
${\cal E}_N$			& -1.7030\hspace{4ex} \\ \hline
${\cal E}  $			&200.9844(5)\hspace{1ex} \\
$24\,\langle\,\pi(\delta^3(r_1)+\delta^3(r_2))\rangle$
				&199.1049\hspace{4ex} \\ \hline
$\Delta{\cal E}_{\rm rel}$	&  1.8795(5)\hspace{1ex} \\
$\Delta{\cal E}_{\rm rec}$	&  -5.5971\hspace{4ex} \\\hline
$\Delta{\cal E}_{\rm th}=\Delta{\cal E}_{\rm rel}+\Delta{\cal E}_{\rm rec}$	
				&  -3.7176(5)\hspace{1ex} \\
$\Delta{\cal E}_{\rm exp}$	& -3.6095(18)
\end{tabular}
\caption{Numerical results for contributions at order $m\,\alpha^6$
to helium hyperfine structure. The factor 24 in the above comes from
Breit correction Eq.(\ref{18}) with $n=1$ times 3/4 from spin algebra times $Z^6/8$.}

\label{table1}
\end{center}
\end{table}


\begin{references}
\bibitem{exception} The experimental value for the  helion magnetic moment is somehow
                    problematic. The reference \cite{mohr} presents values for the screened
                    helion in He$^3$, which involves unknown binding corrections. Therefore we have 
                    decided to follow the reference \cite{hfs} and adopt the value
                    $\mu_h/\mu_p = -0.761\,812\,0(7)$ for the helion--proton
                    magnetic moment ratio.
\end{references}
\end{document}